\documentclass[letterpaper, 10 pt, conference]{ieeeconf}

\IEEEoverridecommandlockouts   

\usepackage[table]{xcolor}
\usepackage{siunitx}
\usepackage{amsmath, amssymb, amsfonts}  
\usepackage[mathscr]{eucal}
\usepackage{cases,enumerate}
\usepackage{mathtools}
\usepackage{blkarray, bigstrut} %

\usepackage{bm}
\usepackage{bbm}
\usepackage{xcolor}
\usepackage{ifpdf}
\usepackage{mleftright}
\usepackage{multicol}
\usepackage{indentfirst}
\usepackage{soul}

\usepackage{graphics} 
\usepackage{graphicx}           
\graphicspath{{./Images/}}

\usepackage{algorithm}
\usepackage{algpseudocode}

\let\labelindent\relax
\usepackage{enumitem}  
\setlist[description]{leftmargin=\parindent}

\usepackage{subfigure}

\usepackage{ifpdf}

\usepackage{theorem}
\newtheorem{theorem}{Theorem}
\newtheorem{definition}{Definition}

\newtheorem{lemma}{Lemma}

\newtheorem{assumption}{Assumption}

\newtheorem{problem}{Problem}

\newcommand{\V}{\mathcal{V}}
\newcommand{\U}{\mathcal{U}}

{\theorembodyfont{\upshape}
	\newtheorem{remark}{Remark}
	\newtheorem{example}{Example}
	
}

\usepackage{stackengine}
\newcommand\barbelow[1]{\stackunder[1.2pt]{$#1$}{\rule{.8ex}{.075ex}}}

\author{Lingfei Wang$^1$, Johan Swartling Sennhed$^2$, David Ohlin$^1$, Emma Tegling$^1$
\thanks{This work is partially funded by Wallenberg AI, Autonomous Systems and Software Program (WASP) as well as the Swedish Research Council through Grant 2025-06419.}
	\thanks{$^{1}$L. Wang, D. Ohlin and E. Tegling are with the Department of Automatic Control and the ELLIIT Strategic Research Area at Lund University, Lund, Sweden. Email:
		{\tt\footnotesize \{lingfei.wang, david.ohlin, emma.tegling\} @control.lth.se}}%
	\thanks{$^{2}$Johan Swartling Sennhed is a Student with the Faculty of Engineering, Lund University, Lund, Sweden. Email:
		{\tt\footnotesize jo1655sw-s@student.lu.se}}%
}

\begin{document}
\title{\LARGE \bf	
	Intransigence or compromise: concatenated opinion dynamics with voting feedback}
\maketitle
	
\begin{abstract}
In this paper, we study opinion evolution in a sequence of Friedkin–Johnsen (FJ) models, concatenated so that the final opinions of each model serve as the initial opinions of the next. Each FJ model is followed by a vote, and each agent becomes partially or fully stubborn in the next model, depending on how close its opinion is to the voting result. With the state-dependent switching of stubbornness, the dynamics are not guaranteed to be contractive. We show that, over any fixed strongly connected network, the opinions always converge. A sufficient condition on the initial opinions is proposed, under which consensus is reached exponentially fast. For the complete graph, we further show that, if the agents' stubbornness levels do not differ too much, an agent that is partially stubborn in one FJ model can never become fully stubborn in a subsequent one: once it has started compromising, it keeps doing so. 
\end{abstract}

\section{Introduction}

Understanding how opinions evolve across multiple issues is an important problem in opinion dynamics. Individuals hold opinions on a range of interconnected issues, and a change in their views on one issue may influence their positions on others. In multidimensional opinion models, the interdependencies are commonly represented by logic matrices, so that the evolution of each individual's opinion depends on both interpersonal influence on the social network and the logical correlation between issues \cite{parsegov2016novel,friedkin2016network}. The multi-layer structure of opinions can also be extended to study other coupled social processes, for instance the coevolution of opinions and decisions \cite{zino2020two,davidson2026modelling}, and the interplay between epidemic spreading and opinion evolution~\cite{she2022networked,peng2021multilayer}.

A complementary line of research, which motivates this paper, considers sequential discussions with opinion evolution occurring on two time scales: opinions evolve within each discussion, while model parameters are updated between discussions. In \cite{jia2015opinion}, a self-appraisal mechanism is introduced into issue-dependent DeGroot models \cite{degroot1974reaching}. Within each issue, agents update their opinions as weighted averages of their neighbours' opinions, while their self-weights are updated between issues according to their influence on the consensus value reached for the preceding issue. The self-appraisal mechanism has been further studied for the Friedkin--Johnsen (FJ) model \cite{mirtabatabaei2014reflected}, which considers agents' attachment to their initial opinions, i.e., stubbornness. Sequential opinion dynamics can also describe repeated negotiations where path-dependence occurs, i.e., the outcome of one discussion shapes the starting point of the next. This is modeled by the concatenated FJ dynamics \cite{TianAutomatica2018,wang2022consensus}, where each discussion is described by an FJ model, and the final opinions every preceding model serve as the initial conditions of the next. As a concrete application, the concatenated FJ model framework has been applied to study consensus formation in the United Nations climate change negotiations culminating in the 2015 Paris Agreement \cite{bernardo2021achieving}.

In complex negotiations, voting is a common mechanism to conclude a discussion round, and its outcome can influence participants' willingness to compromise in subsequent rounds. However, feedback from voting outcomes to opinion dynamics has received relatively limited attention, though introduced in previous work~\cite{ohlin2022achieving}. This paper studies a concatenated FJ model with such voting feedback. Once opinions converge within an FJ event, a vote takes place, with its outcome represented by the median of the final opinions. Each agent's stubbornness in the next event depends on the distance between its own opinion and the voting outcome. An agent becomes intransigent (i.e., fully stubborn), if this distance exceeds a common acceptance threshold; otherwise, it compromises by becoming partially stubborn. Unlike \cite{ohlin2022achieving}, where agents' stubbornness changes monotonically across events, our model allows switching in both directions between full and partial stubbornness. This captures the idea that an agent's willingness to negotiate depends on how closely the collective outcome aligns with its own position.

Our theoretical analysis characterizes the asymptotic behavior of this model. The main analytical challenge is that the voting outcome depends on the entire opinion profile, inducing a nonlinear coupling between opinions and stubbornness. Moreover, the presence of fully stubborn agents means that the dynamics need not be contractive. Nevertheless, we establish convergence of the opinion vector over any fixed strongly connected network, for every nonnegative acceptance threshold (Theorem~\ref{th:convergence}). We further show that consensus is guaranteed when the acceptance threshold is at least half the difference between the largest and smallest initial opinions (Theorem~\ref{th:consensus}). Finally, for complete graphs with uniform influence weights, we prove that, under the same threshold condition and a suitable bound on the heterogeneity of agents' partial stubbornness levels, the set of compromising agents is nondecreasing across discussion events (Theorem~\ref{th:complete_mono}). Thus, once an agent compromises, it continues to do so in all subsequent events.

The remainder of the paper is organized as follows. Section~\ref{sec:preliminary} introduces the necessary preliminaries, presents the model, and formulates the problem. Section~\ref{sec:results} establishes the main theoretical results. Section~\ref{sec:conclusion} concludes the paper. Some preliminary results have appeared in the bachelor thesis \cite{swartling2024dynamic}. It includes Theorem \ref{th:consensus} and the result for odd number of agents in Theorem \ref{th:complete_mono}, with all the proofs redone in this paper. Theorem \ref{th:convergence} and the result for an even number of agents in Theorem \ref{th:complete_mono} are new. 

\section{Preliminaries and problem formulation}\label{sec:preliminary}

\subsection{Notations}

The set of real numbers and nonnegative integers are denoted $\mathbb{R}$ and $\mathbb{N}$, respectively. Given an integer $n>0$, $[n]$ represents the set $\{1,2,\dots, n\}$. All vectors are real column vectors and are denoted by bold lowercase letters $\mathbf{x},\mathbf{y},\dots$. The $i$-th entry of $\mathbf{x}$ is denoted by $x_i$ or $[\mathbf{x}]_i$, with lowercase letters used for scalars. Matrices are denoted by capital letters such as $A,B,\dots$, with entries $[A]_{ij}$ (also denoted by $A_{ij}$ or $a_{ij}$ when no confusion arises). For a matrix $A\in\mathbb{R}^{m\times n}$, $\rho(A)$ represents its spectral radius if $m=n$; given $\mathcal{C}_1\subseteq [m]$ and $\mathcal{C}_2\subseteq [n]$, $A_{\mathcal{C}_1\mathcal{C}_2}$ represents the submatrix of $A$ with row indexes in $\mathcal{C}_1$ and column indexes in $\mathcal{C}_2$. Given a vector $\mathbf{x}\in\mathbb{R}^n$ and a set $\mathcal{C}$, $[\mathbf{x}]_{\mathcal{C}}\subseteq [n]$ is the subvector of $\mathbf{x}$ indexed in $\mathcal{C}$. The identity matrix and the vector of ones are denoted by $I_n$ and $\mathbf{1}_n$, respectively, with dimensions omitted when clear from the context. Given a set of real numbers $\{a_1,\dots, a_n\}$, $\mathrm{diag}(a_1,\dots,a_n)$ denotes the diagonal matrix with diagonal entries $a_1,\dots,a_n$. Sorting $a_1, \dots, a_n$ in an nondecreasing order, $a_{(1)}, \dots, a_{(n)}$, the median of these numbers is denoted by $\mathrm{Median}(a_1,\dots, a_n)$, and defined as 
\[\mathrm{Median}(a_1,\dots, a_n):=\frac{a_{(\lfloor\frac{ n }{2}\rfloor )}+a_{(\lceil\frac{ n }{2}\rceil )}}{2}. \] 

\subsection{Basic notions from graph theory}
	
	Given $\mathcal{V}=[n]$, consider a network with nodes (agents) indexed in $\mathcal{V}$. The network is represented by a directed graph $\mathcal{G}=(\mathcal{V},\mathcal{E})$, where $\mathcal{E}$ is a set of ordered pairs of nodes and $(j,i)\in\mathcal{E}$ represents a link from node $j$ to node $i$. A directed {\em path} is a concatenation of directed links of $\mathcal{E}$.
	We say that node $i$ is connected to node $j$ if there is a directed path from $j$ to $i$. The graph $\mathcal{G}$ is called {\em strongly connected} if any two nodes are connected to each other. 
	
	A matrix is called {\em (row) stochastic} if it is nonnegative and its rows sum to 1.
	Given a stochastic matrix $ {A=[a_{ij}] \in \mathbb{R}^{m \times m }} $, its digraph is denoted as $ \mathcal{G}_A = (\mathcal{M}, \, \mathcal{E}_{\mathcal{M}}) $, where $\mathcal{M}, \mathcal{E}_{\mathcal{M}} $ are the sets of nodes and edges respectively, with $|\mathcal{M}|=m$ and $ (i, \, j ) \in \mathcal{E}_{\mathcal{M}} $ if and only if $ a_{ji} >0$. 

\subsection{The Friedkin-Johnsen (FJ) model}
	
	The Friedkin-Johnsen (FJ) model is a DeGroot-like model for opinion dynamics in which some agents behave stubbornly, in the sense  that they defend their positions while discussing with the other agents \cite{friedkin1999influence}.  If $n$ agents participate to a discussion, the FJ model has the following structure:
	\begin{equation}
	\mathbf{y}(t+1)= (I -\Theta)W \mathbf{y}(t) + \Theta \mathbf{y}(0), \quad t=0,\, 1, \ldots
	\label{eq:FJ1}
	\end{equation}
	where $\mathbf{y} $ is the $n$-dimensional opinion vector (one variable for each agent), $ W $ is a row-stochastic matrix, and $ \Theta = {\rm diag} (\theta_1, \ldots, \theta_m ) $, with $ \theta_i \in [0,1]$, is a diagonal matrix of parameters representing the stubbornness of the $ n$ agents. 
	Stubbornness here means attachment of an agent to its own opinion, represented by the initial condition $ \mathbf{y}(0) $ at the beginning of the discussion ($ \theta_i =0 $ means agent $ i$ is not at all stubborn, $ \theta_i =1 $ means a totally stubborn agent).  
	
	Denote by $ \mathcal{G}_W $ the $n$-node graph associated to $W$. The following Lemma holds \cite{wang2022consensus}.

	\begin{lemma}\label{le:1}
		\label{lemma:DT1}
		Assume $ \theta_i \in (0,1]$ for all $i= 1, \ldots, n$, and $ \theta_i >0$ for at least one $i$.
		If $\mathcal{G}_W $ is strongly connected, then 
		\begin{itemize}
			\item[(a)] $(I - \Theta) W$ is Schur stable, i.e. $\rho ( (I - \Theta) W)<1$,
			\item[(b)]  the matrix $V =(I- (I - \Theta) W)^{-1}\Theta $ is stochastic,
			\item[(c)]  $\mathbf{y}(\infty)=\lim _{t\to +\infty} \mathbf{y}(t)= V\mathbf{y} (0)$,
			\item[(d)] ${\![\mathbf{y}(\infty)]_i \!\in\!{\rm co}([\mathbf{y}(0)]_1,\dots, [\mathbf{y}(0)]_n)}$ for any ${i=1,\dots, n}$,
            \item[(e)] If $\theta_i\in (0,1), \forall i\in\V$, $V_{ij}>0$ holds for all $i, j\in\V$. 
		\end{itemize}
	\end{lemma}

\subsection{The Concatenated FJ model with voting feedback}

Consider a sequence of discussion events represented by the FJ models, indexed by $ s \in \mathbb{N}$. Let $t\in\mathbb{N}$ be the time index for each FJ model.
	Following \cite{jia2015opinion} and \cite{wang2022consensus}, we use an opinion vector depending on both the event $ s$ and the time $ t $, $ \mathbf{y}(s, t)\in\mathbb{R}^n$, and a dynamics having two time scales, i.e., we think of $s$ as the clock of a sequence of opinion forming processes. The opinion update along the $y$-axis follows the FJ dynamics \eqref{eq:FJ1}: 
    \begin{equation*}
        \mathbf{y}(s, t+1)= (I -\Theta(s))W \mathbf{y}(s, t) + \Theta \mathbf{y}(s, 0), \quad t\in \mathbb{N},
    \end{equation*}
    where $W$ is fixed for all $s\in\mathbb{N}$, and $\Theta(s)=\mathrm{diag}\{\theta_1(s),\dots,\theta_n(s)\}$ is a diagonal matrix consisted by the issue-dependent (positive) stubbornness that will be defined below.
    The following assumption is made throughout the paper. 

    \begin{assumption}\label{assum:str_con}
		The underlying graph associated with the weight matrix $W$ is strongly connected.
	\end{assumption}
    
    Under Assumption \ref{assum:str_con}, Lemma \ref{le:1} guarantees the convergence of each FJ model. The final opinions are regarded as ``votes" of the agents in the discussion, with the median taken as the voting result. Specifically, for $s\geq 0$, given the final opinions of the $s$th FJ model, $y_i(s,\infty):=\lim_{t\to\infty}y(s,t), \forall i\in\mathcal{V}$, the voting result is  
    \[\mu(s):=\mathrm{Median}\{y_i(s,\infty):i\in\V\}. \]
    The distance between self-vote and the voting result determines agent stubbornness in the next FJ model. For each agent $i$, let 
    \[\delta_i(s):=|y_i(s,\infty)-\mu(s)| \]
    be the distance between its own opinion and the voting result~$\mu(s)$,
    its stubbornness at $s+1$ becomes 
	\begin{equation}\label{eq:stub}
	\theta_i(s+1) \, = \, \begin{cases}  \theta_i \in (0,1) & \text{ if  }  \, \delta_i(s)\leq \tau, \\
	   1 &  \text{ if  } \, \delta_i(s)>\tau,
	\end{cases}
	\end{equation}
    where $\tau\in (0,1)$ is a uniform width of voting acceptance range for the agents. We make the convention that $\mu(-1):=\mathrm{Median}\{y_i(0,0):i\in\V\}$, $\delta_i(-1):=|y_i(0,0)-\mu(-1)|, \forall i\in\V$, and each $\theta_i(0)$ is defined by~\eqref{eq:stub} for ${s=-1}$. 
	The FJ models are {\em concatenated} such that the initial condition at event $s+1$ corresponds to the asymptotic values at issue $ s$, that is,
    \[y_i(s+1,0)=y_i(s,\infty), \quad \forall i\in\V, s\in\mathbb{N}.  \]

    \begin{remark}
        With the voting feedback in the opinion model, each agent can only switch between fully stubborn and partially stubborn, corresponding to two different modes in real-world negotiations respectively, i.e., being intransigent or willing to compromise.
    \end{remark}

    Denote $ \mathbf{x}(s)  $ the opinion vector of the entire set of $n$ agents at the end of the $ s$-th discussion, i.e., $x_i(s)=y_i(s,0)$ for all $i\in\V, s\geq 0$. 
	Under Assumption \ref{assum:str_con}, the opinion evolution for the concatenated FJ model is
	\begin{equation}
	\mathbf{x}(s+1) = P(s) \mathbf{x}(s),
	\label{eq:FJ2}
	\end{equation}
	where, by Lemma \ref{le:1},
	\begin{equation}
	P(s) = (I-(I-\Theta(s))W)^{-1}\Theta(s).
	\label{eq:Ps}
	\end{equation}
	The concatenation of $s$ discussions becomes therefore a left product of stochastic matrices:
	\begin{equation*}
	\mathbf{x}(s+1) = P(1:s) \mathbf{x}(0)= P(s) \ldots P(1) P(0) \mathbf{x}(0) 
	\end{equation*}
	with each $P(s) $ having the structure \eqref{eq:Ps}.

    With the voting-based stubbornness \eqref{eq:stub}, the initial opinions in minority might have large influence on the group's final opinions, which is different from the concatenated FJ model without the voting feedback \cite{TianAutomatica2018, wang2022consensus}. This is illustrated in the following example.  

\begin{figure}[t]
    \centering
    \includegraphics[width=\columnwidth]{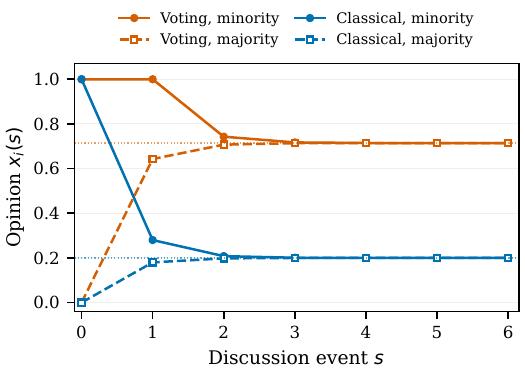}
    \caption{Example~\ref{ex:minority}: event-level opinions with and without voting feedback. The four initially majority agents have identical trajectories. Voting feedback shifts the consensus from $1/5$ to $5/7$.}
    \label{fig:minority}
\end{figure}

\begin{example}\label{ex:minority}
Consider a graph with ${\V=[5]}$ and ${W=\frac{1}{5}\mathbf{1}\mathbf{1}^\top\!}$. In the concatenated FJ model with stubbornness \eqref{eq:stub}, let $\theta_i=\frac{1}{10}$, and in the classical concatenated FJ model without voting feedback \cite{TianAutomatica2018}, stubbornness remains fixed at $\Theta=\frac{1}{10}I$. The vector of initial opinions is $\mathbf{x}(0)=(1,0,0,0,0)^\top$, i.e., agent $1$ has the initial opinion $1$, while the others' are $0$. We run the concatenated FJ model with and without the voting feedback respectively, and plot the opinion evolution in \eqref{fig:minority}. It is clear that with the voting feedback, the initial opinion of agent 1 gains significant influence on the group's final consensus value, compared to the classical concatenated FJ model. More explicitly, by calculation, we have $\lim_{s\to\infty} \mathbf{x}(s)=\frac{1}{5}\mathbf{1}^\top \mathbf{x}(0)$ for the model without voting back, and $\lim_{s\to\infty} \mathbf{x}(s)=(\frac{5}{7}, \frac{1}{14}\mathbf{1}_4^\top) \mathbf{x}(0)$ for the model with voting feedback. The reason for the difference is that, in the concatenated FJ model with voting feedback, if an agent's opinion is far from the voting result, the agent will be fully stubborn and keeps influencing the others. This will eventually steer the group's opinions closer, which in turn gives the agent's initial opinion higher influence on the group's final opinions.
\end{example}

\subsection{Problem formulation}

\begin{definition}\label{def:consensus}
		We say that the concatenated FJ model \eqref{eq:FJ2} achieves consensus if $\lim_{s\to\infty}\mathbf{x}(s)=c\mathbf{1}$ for some $c\in\mathbb{R}$ that depends on $\mathbf{x}(0)$.
	\end{definition}
	
	The problem investigated in this paper is the following.
	\begin{problem}
		\label{prob:formul}
		Consider the concatenated FJ system \eqref{eq:FJ2} where each $ P(s) $ has the structure given in \eqref{eq:stub} and \eqref{eq:Ps}. Characterize the asymptotic bahavior of the model, and find conditions such that \eqref{eq:FJ2} achieves consensus.
	\end{problem}

    
    Under Definition~\ref{def:consensus}, consensus is considered for a given initial opinion vector. This differs from the convention adopted in much of the literature, where consensus is required for arbitrary initial opinions \cite{Blondel2005Convergence,angeli2009convergence,cao2008reaching,Moreau2005Stability,blondel2014decide,xia2018generalized,seneta1979coefficients}. Accordingly, the consensus condition derived below depends on both the acceptance threshold $\tau$ and the initial opinions. The existence of fully stubborn agents means that the dynamics need not be contractive. Consequently, classical consensus analysis tools, including the Lyapunov-based approach proposed in \cite{ohlin2022achieving}, cannot be applied directly.

\section{Main results}\label{sec:results}

Let 
\[\mathcal{M}(s):=\{i\in\V: |x_i(s)-\mu(s-1)|\leq \tau\} \]
denote the set of agents that accept the voting results at event $s$. Define ${\theta_{\min}:=\min_{i\in\V}\theta_i}$, ${\theta_{\max}:=\max_{i\in\V}\theta_i}$ as the minimum and maximum stubbornness levels in the compromising mode, respectively.
For the concatenated FJ model \eqref{eq:FJ2}, the first result is that the opinions will always converge, as shown in the following theorem. 

\begin{theorem}\label{th:convergence}
    Consider the concatenated FJ model \eqref{eq:FJ2}. Let Assumption \ref{assum:str_con} hold. For any $\tau\geq 0$, the opinion vector $\mathbf{x}(s)$ converges as $s$ goes to infinity.
\end{theorem}

\begin{proof}
    We first consider the case $\tau >0$. Define $\bar{x}(s):=\max_{i\in\V}x_i(s)$ and $\barbelow{x}(s):=\min_{i\in\V}x_i(s)$. According to Lemma \ref{le:1}(d) and the concatenation of the model, $\bar{x}(s)$ ($\barbelow{x}(s)$) is monotonically decreasing (increasing). Due to the fact that both $\bar{x}(s)$ and $\barbelow{x}(s)$ are uniformly bounded in $[0,1]$, convergence limits exist, denoted as $x_{\max}$ and $x_{\min}$, respectively for $\bar{x}(s)$ and $\barbelow{x}(s)$. Note that $x_{\max}\geq x_{\min}$. If the equality holds, consensus is achieved, giving the desired conclusion. Thus, we only need to consider $x_{\max}> x_{\min}$. 
    
    We claim that $x_{\max}-x_{\min}\geq \tau$. Otherwise, there exists $\epsilon>0$ and $S_1>0$ such that $\epsilon +x_{\max}-x_{\min}<\tau$ and for all $s\geq S_1$, it holds $x_{\max}\leq \bar{x}(s)<x_{\max}+\frac{\epsilon}{2}$ and $x_{\min}-\frac{\epsilon}{2}<\barbelow{x}(s)\leq x_{\min}$. Therefore, for all $s\geq S_1$, $\bar{x}(s)-\barbelow{x}(s)<\tau$, yielding $i\in\mathcal{M}(s), \forall i\in\V$. Define the Dobrushin ergodicity coefficient as $\sigma(P(s)):=\max_{i,j\in[n]}\sum_{\ell=1}^n|P_{i\ell}(s)-P_{j\ell}(s)|/2$. From Lemma \ref{le:1}(e), it holds $\sigma(P(s))<1, \forall s\geq S_1$. From the property of Dobrushin coefficients \cite{blondel2014decide}, $\lim_{s\to\infty} \sigma(P(s)\dots P(0))\leq \lim_{s\to\infty} \sigma(P(s))\dots\sigma(P(0))=0$, which means that consensus is achieved, contradicting $x_{\max}-x_{\min}>0$. Therefore, $x_{\max}-x_{\min}\geq \tau$. 

    We now prove that $\lim\inf_{s\to \infty}\mathcal{M}^c(s)\not= \emptyset$. Define 
    \[ \beta:=\min_{i\in\V} (1-\theta_i), ~ \, \omega:=\min_{w_{ij}>0} w_{ij}, ~\, \eta:=(\beta \omega)^{n-1}. \]
    Pick an integer $S_2$ such that for all $s\geq S_2$, $\bar{x}(s)\leq x_{\max}+\frac{\theta_{\min}\eta\tau}{2}$ and $\barbelow{x}(s)\geq x_{\min}-\frac{\theta_{\min}\eta\tau}{2}$. From the evolution of concatenated FJ model \eqref{eq:Ps},
    \begin{equation}\label{eq:Von-Neumann}
        P(s)=\sum_{\ell=0}^\infty [(I-\Theta(s))W]^\ell \Theta(s).
    \end{equation}
    Pick $i^\ast, j^\ast\in \V$ such that $x_{i^\ast}(S_2)= \bar{x}(S_2)$ and $x_{j^\ast}(S_2)=\barbelow{x}(S_2)$. 
    \begin{itemize}
        \item \emph{Case} 1: $i^\ast, j^\ast \in\mathcal{M}(S_2)$. Note that for all $i\in\V$, $i\in\mathcal{M}(S_2)$, that is $\theta_i\in(0,1)$. According to the strong connectivity of $\mathcal{G}$, for any $j\in\V$, there exists a path from $i^\ast$ to $j$ with length $L-1\leq n-1$, denoted $i^\ast\to i_1\to\dots\to i_{L-2}\to j$. From \eqref{eq:Von-Neumann},
        \begin{equation}\label{eq:estimate1}
            \begin{aligned}
                [P(S_2)]_{j i^\ast}&\geq \theta_{i^\ast} (1-\theta_{i_1})w_{i_1i^\ast}\dots (1-\theta_{j})w_{j i_{L-2}}\\
                &\geq \theta_{\min} (\beta \omega)^{L-1}\geq \theta_{\min}\eta.
            \end{aligned}
        \end{equation}
        From \eqref{eq:FJ2}, it holds
        \begin{multline}\label{eq:estimate2}
                x_j(S_2+1)\geq P_{ji^\ast}x_{i^\ast}(S_2)+(1-P_{ji^\ast})\barbelow{x}(S_2)\\
                 \geq \theta_{\min} \eta (x_{\min}+\tau)+(1-\theta_{\min}\eta)(x_{\min}-\frac{\theta_{\min}\eta\tau}{2})\\
                 > x_{\min}+\frac{\theta_{\min}\eta\tau}{2},
        \end{multline}
        which gives $\barbelow{x}(S_2+1)>x_{\min}$. This is impossible, since $\barbelow{x}(s)\uparrow x_{\min}$. 
        \item \emph{Case} 2: $i^\ast\in \mathcal{M}^c(S_2), j^\ast \in\mathcal{M}(S_2)$. For any $j\in\mathcal{M}(S_2)$, pick $i_j\in\mathcal{M}^c(S_2)$ such that there exist a path $i_j\to j_1\to\dots \to j_{L_j-2}\to j$, with $L_j\leq n-1$ and $j_k\in\mathcal{M}(S_2), \forall k\in [L_j-2]$. As $i_j\in\mathcal{M}^c(S_2)$, it holds 
        \[x_{i_j}(S_2)-\barbelow{x}(S_2)\geq x_{i_j}(S_2)-\mu(S_2-1)> \tau. \]
        Similarly to \eqref{eq:estimate1}, we can obtain $[P(S_2)]_{ji_j}\geq \eta$, which further gives $x_j(S_2+1)>x_{\min}+\frac{\eta\tau}{2}$. This is again impossible.
        \item \emph{Case} 3: $i^\ast\in \mathcal{M}(S_2), j^\ast \in\mathcal{M}^c(S_2)$. Following similar arguments as for Case 2, this is impossible. 
    \end{itemize}
    From the arguments for Cases 1-3, the only possibility is that $i^\ast, j^\ast\in \mathcal{M}^c(S_2)$, which means that $x_{i^\ast}(S_2+1)=x_{i^\ast}(S_2)=\bar{x}(S_2+1)$ and $x_{j^\ast}(S_2+1)=x_{j^\ast}(S_2)=\barbelow{x}(S_2+1)$. The arguments can then be iterated, that is, $i^\ast, j^\ast \in \lim\inf_{s\to \infty} \mathcal{M}^c(s)$.  

    Let $\mathcal{U}=\lim\inf_{s\to \infty} \mathcal{M}^c(s)$ and $\bar{\mathcal{U}}:=\mathcal{U}^c$. If $\bar{U}=\emptyset$, opinions stay unchanged once all of the agents become fully stubborn. Therefore, we only need to consider the case that $\bar{\U}\not=\emptyset$. From the definition, there exists $S_3>0$ such that for all $s\geq S_3$, $x_i(s+1)=x_i(s), \forall i\in \U$. In the following context, we only consider $s\geq S_3$. Let $\mathbf{h}:=[\mathbf{x}(S_3)]_{\U}$, with $h_i$ representing $[\mathbf{x}(S_3)]_i$ for all $i\in\U$. For the strong connectivity of $\mathcal{G}$, $\rho(W_{\bar{\U}\bar{\U}})<1$, meaning that $I-W_{\bar{\U}\bar{\U}}$ is invertible. We claim that
    \begin{equation}\label{eq:limit}
        \lim_{s\to \infty} [\mathbf{x}(s)]_{\bar{\U}}= (I-W_{\bar{\U}\bar{\U}})^{-1}W_{\bar{\U}\U}\mathbf{h}=:\mathbf{z}.
    \end{equation}
    Note that
    \begin{equation}\label{eq:z_form1}
        \begin{aligned}
            z_i=\sum_{j\in\bar{\U}}w_{ij}z_j+\sum_{j\in\U}w_{ij}h_j, \quad \forall i\in\bar{\U}.
        \end{aligned}
    \end{equation}
    For any $i\in \bar{\U}$, from \eqref{eq:FJ2},
       \begin{multline}\label{eq:z_form2}
                  x_i(s+1)=\theta_i(s) x_i(s)\\
            +(1-\theta_i(s))(\sum_{j\in \bar{\U}}w_{ij}x_j(s+1)+\sum_{j\in\U}w_{ij}h_j)
        \end{multline}
    Subtracting \eqref{eq:z_form1} from \eqref{eq:z_form2} gives the following.
    \begin{equation}\label{eq:error}
            e_i(s+1)\!=\!\theta_i(s) e_i(s) + (1-\theta_i(s))\!\sum_{j\in\bar{\U}}w_{ij}e_j(s+1).
    \end{equation}
    where $e_i(s):=x_i(s)-z_i$. From \eqref{eq:error}, for all $i\in\V$,
    \begin{equation}\label{eq:error2}
        \begin{aligned}
            |e_i(s+1)|\leq \theta_i(s)|e_i(s)|+(1-\theta_i(s))\sum_{j\in\bar{\U}}w_{ij}|e_j(s+1)|.
        \end{aligned}
    \end{equation} 
    Define $r_i=\limsup_{s\to\infty}|e_i(s)|$. Note that for any $s$ such that $i\in\mathcal{M}^c(s)$, it holds $|e_i(s)|=|e_i(s+1)|$, therefore 
    \begin{equation*}
        \begin{aligned}
            r_i=\limsup_{\substack{i\in\mathcal{M}(s)\\ s\to\infty}} |e_i(s+1)|=\limsup_{\substack{i\in\mathcal{M}(s)\\ s\to\infty}} |e_i(s)|.
        \end{aligned}
    \end{equation*}
    Taking the upper limits of the terms at both sides of \eqref{eq:error2},
    \begin{equation*}
        \begin{aligned}
            r_i&\leq \theta_i r_i +\limsup_{\substack{i\in\mathcal{M}(s)\\ s\to\infty}}(1-\theta_i(s))\sum_{j\in\bar{\U}}w_{ij}|e_j(s+1)|\\
            &\leq \theta_i r_i +(1-\theta_i)\sum_{j\in\bar{\U}}w_{ij}r_j,
        \end{aligned}
    \end{equation*}
    which yields $r_i\leq \sum_{j\in\bar{\U}}w_{ij}r_j, \forall i\in\bar{\U}$. Note that $r_i\geq 0, \forall i\in\bar{\U}$. Let $\bar{r}:=\max_{i\in\bar{\U}}r_i$. If $\bar{r}>0$, for all $i\in\bar{\U}$ with $r_i=\bar{r}$, it holds 
    \[\bar{r}\leq \sum_{j\in\bar{\U}}w_{ij}r_j \leq \sum_{j\in\bar{\U}}w_{ij}\bar{r}\leq \bar{r}. \]
    The equality hold only when $\mathcal{N}_i\subseteq \bar{\U}$ and $r_j=\bar{r}, \forall j\in \mathcal{N}_i$. The process can be iterated for $j\in\mathcal{N}_i$. However, due to the strong connectivity, there must exist $i'\in\bar{\U}$ such that $\mathcal{N}_{i'}\cap \U\not=\emptyset$. The contradiction is then obtained. Therefore, it must be $\bar{r}=0$, which proves \eqref{eq:limit}. 

    Now we consider the case that $\tau=0$. For this case, only the agents with median opinion $\mu(s)$ belong to $\mathcal{M}(s)$. If $x_i(0)=x_j(0)$ for all $i,j\in\V$, consensus is achieved from the beginning. Otherwise, there exists an agent $i'$ with the maximum or minimum initial opinion that belongs to $\mathcal{M}^c(0)$. Since $\theta_{i}>0$ for all $i\in\V$, if $x_i(0)\not= x_{i'}(0)$, it must be $x_i(1)\not=x_{i'}(1)=x_{i'}(0)$, which means that $i'\in\mathcal{M}^c(1)$ and $x_{i'}(1)$ is still an extreme opinion. The argument can be iterated, giving that $i'\in\lim\inf_{s\to\infty} \mathcal{M}^c(s)$. We can then apply the same arguments as above, and the proof is completed.
\end{proof}

In the proof above, Eq. \eqref{eq:limit} provides an explicit form of the final opinion vector in terms of the opinions of the persistently intransigent agents (i.e., the agents in $\mathcal{U}$). In general,  the set $\mathcal{U}$ is difficult to characterize from the initial opinions, since the modes of agents (i.e., intransigence or compromise) depend on the opinions of the others and may change inconsistently. Nevertheless, if $\mathcal{M}(s)$ is monotone, as for the case in the following Theorem \ref{th:complete_mono}, $\mathcal{U}$ becomes more predictable since for any $s>0$, $\mathcal{U}$ is a subset of $\cap_{\ell\geq s}\mathcal{M}^c(\ell)$, which is also monotone w.r.t $s$.

\begin{remark}
    From \eqref{eq:limit} in the proof, if consensus is not achieved, all the opinions in the CFJ model \eqref{eq:FJ2} will converge to the convex hull spanned by the fully stubborn opinions, i.e., the opinions of the agents in $\mathcal{U}$. In particular, there exists fully stubborn agents $i',j'\in\mathcal{U}$ with $x_{i'}(\infty)<x_{j'}(\infty)$, such that the opinions of all compromised agents (i.e., $\mathcal{U}^c$) sit in $(x_{i'}(\infty),x_{j'}(\infty))$. For $i'$ and $j'$ to be fully stubborn, it must be $x_{i'}(\infty)<\mu(\infty)-\tau$ and $x_{j'}(\infty)>\mu(\infty)+\tau$, which gives $\tau<\frac{\max_{i,j\in\V}|x_{i}(s)-x_{j}(s)|}{2}$ for large enough $s$. On the other hand, if $\tau\geq\frac{\max_{i,j\in\V}|x_{i}(s)-x_{j}(s)|}{2}$ for some $s\geq 0$, consensus will be achieved, as shown in the following Theorem \ref{th:consensus}.
\end{remark}

Before introducing Theorem \ref{th:consensus}, define 
\[d(s):=\max_{i,j\in\V} |x_i(s)-x_j(s)| \]
as the width of the smallest interval that contains all the opinions $x_1(s),\dots, x_n(s)$. 

\begin{theorem}\label{th:consensus}
    Consider the concatenated FJ model \eqref{eq:FJ2}. Let Assumption \ref{assum:str_con} hold. If $\tau\geq \frac{d(0)}{2}$, consensus will be achieved as $s$ goes to infinity.
\end{theorem}

\begin{proof}
    The notations in the proof of Theorem \ref{th:convergence} are followed. From Lemma \ref{le:1} and the model dynamics \eqref{eq:FJ2}, for all $s\geq 0$, $x_i(s+1)\in[\barbelow{x}(s),\bar{x}(s)], \forall i\in\V$, which means $d(s+1)\leq d(s)$, and consequently, $\tau\geq \frac{d(s)}{2}$. Therefore, $i\in\mathcal{M}(s)$ holds either for all $i$ with $x_i(s)\leq \mu(s-1)$ or for all $i$ with $x_i(s)\geq \mu(s-1)$.  
    \begin{itemize}
        \item \emph{Case} $1$: $i\in\mathcal{M}(s), \forall i\in\V$. From \eqref{eq:estimate1}, $[P(s)]_{ij}\geq \theta_{\min} \eta$ holds for all $i,j\in\V$. The Dobrushin ergodicity coefficient is $\sigma(P(s))\leq 1-n\theta_{\min}\eta$. From the property of the coefficient (see Eq. (1) in \cite{gaubert2015dobrushin}), 
        \begin{equation}\label{eq:shrink1}
            d(s+1)\leq \sigma(P(s))d(s)\leq (1-n\theta_{\min}\eta)d(s).
        \end{equation}
        \item \emph{Case} $2$: $\exists i^\ast\in\V$ such that $i^\ast\in \mathcal{M}^c(s)$. Without loss of generality, assume that $x_{i^\ast}(s)=\bar{x}(s)$. Through the same analysis as for Case 2 in the proof of Theorem \ref{th:convergence}, for any $j\in\mathcal{M}(s)$, there exists $i_j\in\mathcal{M}^c(s)$ such that $[P(s)]_{ji_j}\geq \eta$, which gives $x_j(s+1)\geq \eta x_{i_j}(s)+(1-\eta)\barbelow{x}(s)\geq \eta (\barbelow{x}(s)+\tau)+(1-\eta)\barbelow{x}(s)\geq \barbelow{x}(s)+\frac{\eta d(s)}{2}$. For $j\in\mathcal{M}(s)$, it holds $x_j(s+1)=x_j(s)\geq \barbelow{x}(s)+\tau\geq \barbelow{x}(s)+\frac{d(s)}{2}$.
        Therefore, 
        \begin{multline} \label{eq:shrink2}
                d(s+1)=\bar{x}(s+1)-\barbelow{x}(s+1)=\bar{x}(s)-\barbelow{x}(s+1)\\
                \leq \bar{x}(s)-\barbelow{x}(s)-\min\{\frac{\eta }{2},\frac{1}{2} \}d(s)\\
                = (1-\min\{\frac{\eta }{2},\frac{1}{2} \})d(s)
        \end{multline}
    \end{itemize}
    Combining \eqref{eq:shrink1} and \eqref{eq:shrink2}, $d(s)$ converges to $0$ exponentially. The proof is then completed.
\end{proof}

Intuitively, $\tau\geq \frac{d(0)}{2}$ means that the agents are sufficiently open from the start. From the proof, this condition requires the compromise of at least half of the agents, with their opinions either above or below the median $\mu(s-1)$. This lets the opinion interval $x_{\min}(s), x_{\max}(s)$ keep shrinking, and leads to a final consensus. The bound $\tau\geq \frac{d(0)}{2}$ cannot be improved for achieving opinion consensus. If $\tau< \frac{d(0)}{2}$, we can always find a graph with $n=2k$ and $\mathbf{x}(0)=(\mathbf{0}_k^\top, \mathbf{1}_k^\top)^\top$ such that consensus will not be achieved, since in this case all the agents will be intransigent from the beginning.

\begin{remark}
    The condition $\tau\geq \frac{d(0)}{2}$ is not necessary for achieving opinion consensus. For instance, consider a graph of $3$ agents, with $\tau=\frac{2}{5}$, $\mathbf{x}(0)=(1,0,0)^\top$, $\theta_1=\theta_2=\theta_3=\frac{1}{10}$, and
    \[W=\left(\begin{array}{ccc}
        \frac{1}{3} & \frac{1}{3} & \frac{1}{3}\\
        \frac{9}{10} & \frac{1}{10} & 0\\
        \frac{9}{10} & 0 & \frac{1}{10}
    \end{array}\right).\] 
    It can be seen that $\tau <\frac{1}{2}=\frac{d(0)}{2}$. However, $\mathbf{x}(1)=(1,\frac{81}{100}, \frac{81}{100})^\top$, giving that $\tau > \frac{d(1)}{2}$, therefore consensus will be finally achieved. 
\end{remark}

As seen from the proof above, the existence of compromising agents is essential for achieving opinion consensus. A natural question then follows: under what circumstances can an agent who has once begun to compromise be guaranteed to never again become fully stubborn? The problem is complicated for general graphs, as the stubbornness of each agent is coupled with the whole opinion vector in a highly nonlinear way. Nevertheless, if the graph is a complete graph with a rank-$1$ weight matrix, a simple condition that keeps the monotonicity of the compromised agent set can be proposed:

\begin{theorem}\label{th:complete_mono}
Consider the concatenated FJ model \eqref{eq:FJ2} with $W=\frac1n\mathbf1\mathbf1^\top $.
If $\tau\ge\frac{d(0)}{2}$ and $\theta_{\max}\leq \frac{1+\theta_{\min}}{2}$, it holds $\mathcal{M}(s)\subseteq\mathcal{M}(s+1), \forall s\in\mathbb{N}$.
\end{theorem}

\begin{proof}
    From the proof of Theorem \ref{th:consensus}, $d(s)\leq 2\tau$ holds for all $s\in\mathbb{N}$. Applying the Sherman-Morrison formula to \eqref{eq:Ps}, we obtain $P(s)=\Theta(s)+\frac{(I-\Theta(s))\mathbf{11}^\top \Theta(s)}{\sum_{i\in\V}\theta_i(s)}$. Therefore, for $s\in\mathbb{N}$,
    \begin{equation}\label{eq:CFJ_complete}
        x_i(s+1)=\theta_i(s)x_i(s)+(1-\theta_i(s))z(s), \quad \forall i\in\V,
    \end{equation}
    where $z(s):=\frac{\sum_{j\in\V}\theta_j(s)x_j(s)}{\sum_{j\in\V}\theta_j(s)}$. We first prove the following claim.
    \\ \hfill{} \\
    \emph{Claim 1}. For $i, j\in\mathcal{M}(s)$, if $|x_i(s)-x_j(s)|\leq \tau$, it holds $|x_i(s+1)-x_j(s+1)|\leq \tau$.
    \\ \hfill{} \\
    From \eqref{eq:CFJ_complete}, it holds
    \begin{equation}\label{eq:claim1_es1}
        \begin{aligned}
            &x_i(s+1)-x_j(s+1)\\
            &=\theta_i(s)(x_i(s)-z(s))-\theta_j(s)(x_j(s)-z(s))\\
            &=(\theta_i(s)-\theta_j(s))(x_i(s)-z(s))+\theta_j(s)(x_i(s)-x_j(s)).
        \end{aligned}
    \end{equation}
    Without loss of generality, assume $\theta_i(s)\geq \theta_j(s)$. Therefore,
    \begin{multline*}
            |x_i(s+1)-x_j(s+1)|\\
            \leq (\theta_i(s)-\theta_j(s))|x_i(s)-z(s)|+\theta_j(s)|x_i(s)-x_j(s)|\\
            \leq (\theta_i(s)-\theta_j(s))2\tau+\theta_j(s)\tau \\
            = (2\theta_i(s)-\theta_j(s))\tau\leq (2\theta_{\max}-\theta_{\min})\tau\leq \tau,
    \end{multline*}
    where the second inequality is from $d(s)\leq 2\tau$, and the last inequality is from the assumption. Claim 1 is then proved. In the remaining proof, for simplicity of notation, denote $x_j^+=x_j(s+1), x_j=x_j(s), \forall j\in\V$, $\mu:=\mu(s-1)$, $\mu^+=\mu(s)$, $z:=z(s)$ and $d:=d(s)$. 

    Now consider an arbitrary $i\in\mathcal{M}(s)$. For any $j\in\V$ with $x_j \geq \mu$, we can prove that $x_j^+\geq x_i^+-\tau$:
    \begin{itemize}
        \item If $x_j\leq x_i$ and $j\in\mathcal{M}(s)$,  it must be $|x_i-x_j|\leq \tau$. By Claim 1, $|x_i^+-x_j^+|\leq \tau$.
        \item If $x_i\leq x_j$ and $j\in\mathcal{M}^c(s)$, it holds  $x_j^+=x_j$, and therefore,
        \begin{multline*}
                x_i^+-x_j^+=\theta_i(s)x_i+(1-\theta_i(s))z-x_j\\
                =\theta_i(s)(x_i-x_j)+(1-\theta_i(s))(z-x_j)\\
                \leq (1-\theta_i(s))(\bar{x}(s)-x_j)\leq \tau,
        \end{multline*}
        where the second inequality is from $x_j>\mu+\tau$ and $d\leq 2\tau$.  
        \item If $x_i\leq x_j$ and $j\in\mathcal{M}(s)$, from \eqref{eq:claim1_es1}
        \begin{multline*}
                x_i^+-x_j^+\leq (\theta_i(s)-\theta_j(s))(x_i-z)\\
                \leq (\theta_{\max}-\theta_{\min})2\tau\leq (2\theta_{\max}-\theta_{\min})\tau\leq \tau.
        \end{multline*}
    \end{itemize}
    Similarly, for any $j\in\V$ with $x_j \leq \mu$,  we can prove that $x_j^+\leq x_i^++\tau$:
    \begin{itemize}
        \item If $x_i\leq x_j$ and $j\in\mathcal{M}(s)$, by Claim 1, $|x_i^+-x_j^+|\leq \tau$.
        \item If $x_i\geq x_j$ and $j\in\mathcal{M}^c(s)$, it holds $x_j^+=x_j$, and $x_i^+-x_j^+=x_i^+-x_j=\theta_i(s)(x_i-x_j)+(1-\theta_i(s))(z-x_j)\geq (1-\theta_i(s))(\barbelow{x}(s)-x_j)\geq -\tau$.
        \item If $x_i\geq x_j$ and $j\in\mathcal{M}(s)$, by \eqref{eq:claim1_es1}, it holds $x_i^+-x_j^+\geq (\theta_i(s)-\theta_j(s))(x_i-z)\geq -(\theta_{\max}-\theta_{\min})2\tau\geq -\tau$.
    \end{itemize}
    Sort $x_j^+, j\in\V$ in nondecreasing order, and index the sorted sequence by $x_{v_\ell}^+, \ell \in\V$. We make the discussion for $n=2k+1$ and $n=2k$ respectively.

    If $n=2k+1$ for some $k\in\mathbb{N}$, it holds $\#\{j:x_j\geq \mu\}\geq k+1$ and $\#\{j:x_j\leq \mu\}\geq k+1$. Therefore, from the conclusions above, it must be
    \[x_{v_{k+1}}^+\geq x_i^+-\tau, \; x_{v_{k+1}}^+\leq x_i^++\tau. \]
    Due to the fact that $\mu^+=x_{v_{k+1}}^+$, it is $|x_i^+-\mu^+|\leq \tau$, i.e., $i\in\mathcal{M}(s+1)$.

    If $n=2k$, it holds $\#\{j:x_j\geq \mu\}\geq k$ and $\#\{j:x_j\leq \mu\}\geq k$. Therefore, 
    \[x_{v_{k+1}}^+\geq x_i^+-\tau, \; x_{v_k}^+\leq x_i^++\tau. \]
    Sort $x_j$ in a nondecreasing order, and denote the reordered sequence as $x_{r_\ell}, \ell\in\V$. Note that $\mu=\frac{x_{r_k}+x_{r_{k+1}}}{2}$. We first prove that $x_i^+\leq \mu^++\tau$. This already holds if $x_i^+\leq x_{v_k}^++\tau$. Therefore, we only need to consider the case $x_i^+> x_{v_k}^++\tau$. From the proof above, $x_i^+\leq x_{r_\ell}^++\tau$ holds for all $\ell\geq k+1$. Therefore, $i\in\mathcal{C}:= \{r_{\ell}:\ell\geq k+1\}$, and by Claim 1, it must be $x_i-x_{r_k}>\tau$.  

    \emph{Step} 1. We prove that for any $u\in \mathcal{C}$, it holds 
    \begin{equation*}
        \Delta_u:=2x_i^+-x_{r_k}^+-x_u^+\leq 2\tau.
    \end{equation*}
    From \eqref{eq:CFJ_complete},
    \begin{equation}\label{eq:even_es2}
        \begin{aligned}
            \Delta_u=2\theta_i(x_i-z)-\theta_{r_k}(x_{r_k}-z)-\theta_u(s)(x_u-z).
        \end{aligned}
    \end{equation}
    \begin{itemize}
        \item For $z\leq x_{r_k}$, from $x_{r_k}\leq x_i$ and $x_{r_k}\leq x_u$,
        \begin{equation*}
            \begin{aligned}
                &\Delta_u\leq 2\theta_{\max}(x_i-z)-\theta_{\min}(x_{r_k}+x_u-2z)\\
                &=2(\theta_{\max}-\theta_{\min})(x_i-z)+\theta_{\min}(2x_i-x_{r_k}-x_u)\\
                &\leq 4(\theta_{\max}-\theta_{\min})\tau+2\theta_{\min}\tau= (4\theta_{\max}-2\theta_{\min})\tau\leq 2\tau,
            \end{aligned}
        \end{equation*}
        where the second inequality is from $|x_i-z|\leq d\leq 2\tau$ and $x_i-\frac{x_{r_k}+x_u}{2}\leq x_i-\mu\leq \tau$.
        \item For $x_{r_k}<z\leq x_u$, from \eqref{eq:even_es2},
        \begin{equation*}
            \begin{aligned}
                &\Delta_u=\theta_i(2x_i-x_{r_k}-x_u)+(\theta_{r_k}-\theta_i)(z-x_{r_k})\\
                &+(\theta_i-\theta_u(s))(x_u-z)\leq \theta_{\max}2\tau\\
                &+(\theta_{\max}-\theta_{\min})(z-x_{r_k})+(\theta_{\max}-\theta_{\min})(x_u-z)\\
                &\leq \theta_{\max}2\tau+(\theta_{\max}-\theta_{\min})(x_u-x_{r_k})\\
                &\leq \theta_{\max}2\tau+(\theta_{\max}-\theta_{\min})2\tau =(4\theta_{\max}-2\theta_{\min})\tau\leq 2\tau,
            \end{aligned}
        \end{equation*}
        where the first inequality is from $x_i-\frac{x_{r_k}+x_u}{2}\leq \tau$, $z-x_{r_k}\leq 2\tau$ and $x_u-z\leq 2\tau$.
        \item For $x_u<z$ and $u\in\mathcal{M}(s)$, if $x_i>z$,
        \begin{equation*}
            \begin{aligned}
                &\Delta_u=2\theta_i(x_i-z)+\theta_{r_k}(z-x_{r_k})+\theta_u(z-x_u)\\
                &\leq 2\theta_{\max}(x_i-z)+\theta_{\max}(z-x_{r_k})+\theta_{\max}(z-x_u)\\
                &=\theta_{\max}(2x_i-x_{r_k}-x_u)\leq 2\theta_{\max}\tau\leq 2\tau,
            \end{aligned}
        \end{equation*}
        and if $x_i\leq z$, from \eqref{eq:even_es2},
        \begin{equation*}
            \begin{aligned}
                &\Delta_u\leq -2\theta_{\min}(z-x_i)+\theta_{\max}(2z-x_{r_k}-x_u)\\
                &=2(\theta_{\max}-\theta_{\min})(z-x_i)+\theta_{\max}(2x_i-x_{r_k}-x_u)\\
                &\leq 2(\theta_{\max}-\theta_{\min})\tau + \theta_{\max}2\tau\leq 2\tau,
            \end{aligned}
        \end{equation*}
        where the second inequality is from $x_i-x_{r_k}>\tau$ and $d\leq 2\tau$, which gives $z-x_i<\tau$. 
        \item For $x_u<z$ and $u\in\mathcal{M}^c(s)$, it holds $x_i<x_u<z$, and $z-x_i\leq \tau$. Therefore,
        \begin{equation*}
            \begin{aligned}
                &\Delta_u\leq  -2\theta_{\min}(z-x_i)+\theta_{\max}(z-x_{r_k})+z-x_u\\
                &< (1-2\theta_{\min})(z-x_i)+\theta_{\max}(z-x_{r_k})\\
                &\leq \min\{1-2\theta_{\min}, 0\}\tau +\theta_{\max}2\tau\leq 2\tau.
            \end{aligned}
        \end{equation*}
    \end{itemize}

    \emph{Step} 2. With $\Delta_u\leq 2\tau$ for all $u\in \mathcal{C}$, we have
    \begin{equation*}
        \begin{aligned}
            2x_i^+-x_{v_k}^+-x_{v_{k+1}}^+\leq 2x_i^+-x_{r_k}^+-\min_{u\in\mathcal{C}} x_u^+\leq 2\tau.
        \end{aligned}
    \end{equation*}
    That is, $x_i^+\leq \mu^++\tau$. 

    Note that the analysis above does not rely on specific values of $x_j$, and if the opinion of each agent $j\in\V$ in the CFJ model (at time $s$) is $\tilde{x}_j=-x_j$, the arguments still hold. In this case, the opinion of agent $j$ at time $s+1$ is $\tilde{x}_j^+=-x_j^+$. Let $\tilde{\mu}^+:=\mathrm{Median}\{\tilde{x}_j^+:j\in\V\}=-\mu^+$. It then holds $\tilde{x}_i^+\leq \tilde{\mu}^++\tau$, i.e., $-x_i^+\leq -\mu^++\tau$, which gives $x_i^+\geq \mu^+-\tau$. 

    Combining all the arguments above for $n=2k$, it holds $i\in\mathcal{M}(s+1)$. The proof is then completed.

\end{proof}


The condition $\theta_{\max}\leq \frac{1+\theta_{\min}}{2}$ implies that the stubbornness of the agents do not differ much, thereby limiting differences in how compromising agents respond to social influence during an FJ event. In particular, it prevents some agents from moving disproportionately towards intransigent opinions while others change relatively little, ensuring that previously compromising agents remain within the acceptance interval around the updated opinion median.

\section{Conclusion}\label{sec:conclusion}

This paper studied a concatenated FJ model with voting feedback, in which agents switch between fully and partially stubborn modes according to their distance from the opinion median. Despite this binary switching and the possible absence of contraction, we established convergence of the opinion vector over any fixed strongly connected network. We also showed that an acceptance threshold at least half the initial opinion diameter guarantees exponentially fast consensus. For complete graphs with uniform weights, this threshold condition, together with a bound on the heterogeneity of the partial stubbornness levels, ensures that the set of compromised agents is nondecreasing: once an agent compromises, it continues to do so in subsequent discussions. These results illustrated how the interplay between the acceptance threshold and distribution of the initial opinions influences the consensus forming process. Future work includes characterizing consensus under smaller acceptance thresholds, and extending the analysis to time-varying networks and alternative voting rules.

\bibliographystyle{abbrv}
\bibliography{ref}

@article{friedkin1999influence,
	title={Influence networks and opinion change},
	author={Friedkin, Noah E and Johnsen, Eugene C},
	journal={Advances in Group Processes},
	volume={16},
	number={1},
	pages={1--29},
	year={1999}
}

@article{wang2022consensus,
  title={Consensus in concatenated opinion dynamics with stubborn agents},
  author={Wang, Lingfei and Bernardo, Carmela and Hong, Yiguang and Vasca, Francesco and Shi, Guodong and Altafini, Claudio},
  journal={IEEE Transactions on Automatic Control},
  volume={68},
  number={7},
  pages={4008--4023},
  year={2022},
  publisher={IEEE}
}

@article{jia2015opinion,
	title={Opinion dynamics and the evolution of social power in influence networks},
	author={Jia, Peng and Mirtabatabaei, Anahita and Friedkin, Noah E and Bullo, Francesco},
	journal={SIAM review},
	volume={57},
	number={3},
	pages={367--397},
	year={2015},
	publisher={SIAM}
}

@article{TianAutomatica2018,
	title={Opinion dynamics in social networks with stubborn agents: An issue-based perspective},
	author={Tian, Ye and Wang, Long},
	journal={Automatica},
	volume={96},
	pages={213--223},
	year={2018},
	publisher={Elsevier}
}

@inproceedings{ohlin2022achieving,
  title={Achieving consensus in networks of increasingly stubborn voters},
  author={Ohlin, David and Bencherki, Fethi and Tegling, Emma},
  booktitle={2022 IEEE 61st Conference on Decision and Control (CDC)},
  pages={3531--3537},
  year={2022},
  organization={IEEE}
}

@INPROCEEDINGS{Blondel2005Convergence,
  author={V. D. {Blondel} and J. M. {Hendrickx} and A. {Olshevsky} and J. N. {Tsitsiklis}},
  booktitle={Proceedings of the 44th IEEE Conference on Decision and Control}, 
  title={Convergence in Multiagent Coordination, Consensus, and Flocking}, 
  year={2005},
  volume={},
  number={},
  pages={2996-3000},
  doi={10.1109/CDC.2005.1582620}}

@article{angeli2009convergence,
author = {Angeli, David and Bliman, Pierre-Alexandre},
title = {Convergence Speed of Unsteady Distributed Consensus: Decay Estimate Along the Settling Spanning-Trees},
journal = {SIAM Journal on Control and Optimization},
volume = {48},
number = {1},
pages = {1-32},
year = {2009},
doi = {10.1137/060673527},
}

@article{cao2008reaching,
author = {Cao, Ming and Morse, A. Stephen and Anderson, Brian D. O.},
title = {Reaching a Consensus in a Dynamically Changing Environment: A Graphical Approach},
journal = {SIAM Journal on Control and Optimization},
volume = {47},
number = {2},
pages = {575-600},
year = {2008},
doi = {10.1137/060657005},
}

@ARTICLE{Moreau2005Stability,
author={Moreau, L.},
journal={IEEE Transactions on Automatic Control}, 
title={Stability of multiagent systems with time-dependent communication links},
year={2005},
month={feb.},
volume={50},
number={2},
pages={ 169 - 182},
doi={10.1109/TAC.2004.841888},
ISSN={0018-9286},}

@article{blondel2014decide,
	title={How to decide consensus? A combinatorial necessary and sufficient condition and a proof that consensus is decidable but NP-hard},
	author={Blondel, Vincent D and Olshevsky, Alex},
	journal={SIAM Journal on Control and Optimization},
	volume={52},
	number={5},
	pages={2707--2726},
	year={2014},
	publisher={SIAM}
}

@article{xia2018generalized,
	title={Generalized sarymsakov matrices},
	author={Xia, Weiguo and Liu, Ji and Cao, Ming and Johansson, Karl Henrik and Ba{\c{s}}ar, Tamer},
	journal={IEEE Transactions on Automatic Control},
	volume={64},
	number={8},
	pages={3085--3100},
	year={2018},
	publisher={IEEE}
}

@article{seneta1979coefficients,
	title={Coefficients of ergodicity: structure and applications},
	author={Seneta, Eugene},
	journal={Advances in applied probability},
	volume={11},
	number={3},
	pages={576--590},
	year={1979},
	publisher={Cambridge University Press}
}

@article{gaubert2015dobrushin,
  title={Dobrushin’s ergodicity coefficient for Markov operators on cones},
  author={Gaubert, St{\'e}phane and Qu, Zheng},
  journal={Integral Equations and Operator Theory},
  volume={81},
  number={1},
  pages={127--150},
  year={2015},
  publisher={Springer}
}

@article{she2022networked,
  title={On a networked SIS epidemic model with cooperative and antagonistic opinion dynamics},
  author={She, Baike and Liu, Ji and Sundaram, Shreyas and Par{\'e}, Philip E},
  journal={IEEE Transactions on Control of Network Systems},
  volume={9},
  number={3},
  pages={1154--1165},
  year={2022},
  publisher={IEEE}
}

@article{peng2021multilayer,
  title={A multilayer network model of the coevolution of the spread of a disease and competing opinions},
  author={Peng, Kaiyan and Lu, Zheng and Lin, Vanessa and Lindstrom, Michael R and Parkinson, Christian and Wang, Chuntian and Bertozzi, Andrea L and Porter, Mason A},
  journal={Mathematical Models and Methods in Applied Sciences},
  volume={31},
  number={12},
  pages={2455--2494},
  year={2021},
  publisher={World Scientific}
}

@article{zino2020two,
  title={A two-layer model for coevolving opinion dynamics and collective decision-making in complex social systems},
  author={Zino, Lorenzo and Ye, Mengbin and Cao, Ming},
  journal={Chaos: An Interdisciplinary Journal of Nonlinear Science},
  volume={30},
  number={8},
  year={2020},
  publisher={AIP Publishing}
}

@article{davidson2026modelling,
  title={Modelling the coevolution of opinion dynamics and decision making in social dilemmas},
  author={Davidson, Ella C and Zino, Lorenzo and Cao, Ming and Ye, Mengbin},
  journal={arXiv preprint arXiv:2604.08840},
  year={2026}
}

@article{degroot1974reaching,
  title={Reaching a consensus},
  author={DeGroot, Morris H},
  journal={Journal of the American Statistical association},
  volume={69},
  number={345},
  pages={118--121},
  year={1974},
  publisher={Taylor \& Francis}
}

@inproceedings{mirtabatabaei2014reflected,
  title={On the reflected appraisals dynamics of influence networks with stubborn agents},
  author={Mirtabatabaei, Anahita and Jia, Peng and Friedkin, Noah E and Bullo, Francesco},
  booktitle={2014 American Control Conference},
  pages={3978--3983},
  year={2014},
  organization={IEEE}
}

@article{bernardo2021achieving,
  title={Achieving consensus in multilateral international negotiations: The case study of the 2015 Paris Agreement on climate change},
  author={Bernardo, Carmela and Wang, Lingfei and Vasca, Francesco and Hong, Yiguang and Shi, Guodong and Altafini, Claudio},
  journal={Science Advances},
  volume={7},
  number={51},
  pages={eabg8068},
  year={2021},
  publisher={American Association for the Advancement of Science}
}

@article{parsegov2016novel,
  title={Novel multidimensional models of opinion dynamics in social networks},
  author={Parsegov, Sergey E and Proskurnikov, Anton V and Tempo, Roberto and Friedkin, Noah E},
  journal={IEEE Transactions on Automatic Control},
  volume={62},
  number={5},
  pages={2270--2285},
  year={2016},
  publisher={IEEE}
}

@article{friedkin2016network,
  title={Network science on belief system dynamics under logic constraints},
  author={Friedkin, Noah E and Proskurnikov, Anton V and Tempo, Roberto and Parsegov, Sergey E},
  journal={Science},
  volume={354},
  number={6310},
  pages={321--326},
  year={2016},
  publisher={American Association for the Advancement of Science}
}

@article{swartling2024dynamic,
  title={On dynamic stubbornness in the concatenated {Friedkin-Johnsen} model},
  author={Swartling Sennhed, Johan},
  year={2024}
}
\end{document}